\documentclass{article}

\usepackage{graphicx,tikz}
\usepackage{fullpage}
\usepackage{amsthm}
\usepackage{amssymb}
\usepackage{amsmath}
\usepackage{algorithm}
\usepackage{algorithmicx}
\usepackage{algpseudocode}
\usepackage{booktabs}

\begin{document}

\title{\Large\bf An Asymptotically Tighter Bound on Sampling for Frequent Itemsets Mining}
\author{Shiyu Ji, Kun Wan\\ \{shiyu,kun\}@cs.ucsb.edu}
\date{}
\maketitle

\newtheorem{definition}{Definition}
\theoremstyle{definition}
\newtheorem{theorem}{Theorem}
\theoremstyle{plain}
\newtheorem{lemma}{Lemma}
\theoremstyle{plain}
\newtheorem{corollary}{Corollary}
\theoremstyle{plain}

\begin{abstract}
In this paper we present a new error bound on sampling algorithms for frequent itemsets mining. We show that the new bound is asymptotically tighter than the state-of-art bounds, i.e., given the chosen samples, for small enough error probability, the new error bound is roughly half of the existing bounds. Based on the new bound, we give a new approximation algorithm, which is much simpler compared to the existing approximation algorithms, but can also guarantee the worst approximation error with precomputed sample size. We also give an algorithm which can approximate the top-$k$ frequent itemsets with high accuracy and efficiency.
\end{abstract}

\section{Introduction}
Frequent Itemsets (FI) mining has been popular in research recently \cite{AIS93, HCX07, RU15}. The goal of FI mining is to find out the items that most frequently appear in the observed transactions, e.g., the researchers who are the most prolific in writing papers with others, the patterns that appear frequently in long pieces of genetic code, etc. 

In the era of big data, to compute the exact frequencies can be very time consuming. Thus in many cases approximate values are also acceptable \cite{AIS93,PCY95,FSG99,HCX07,LRU14,RU15}.
For FI mining in large scale transactional datasets, we often take samplings on the transactions, and compute the frequencies of the itemsets among the sampled transactions as approximate results of their true frequencies among all the transactions. Usually the sampling size is much less than the scale of all the transactions, and the approximations can achieve acceptable precision. Also in reality we often only want to know the most frequent itemsets without the need of their actual frequencies, and there are already many works \cite{AIS93,PCY95,FSG99,HCX07,LRU14} on this area. Thus FI approximation can be useful in practice.

The state-of-art progressive sampling based FI approximation algorithms \cite{RU15} need an upper bound of the approximation error for the worst case, i.e., the maximum error the algorithm can generate among all the items. The algorithms keep taking new samples until the upper bound is less than the acceptable threshold. Hence how to bound the maximum error as tightly as possible is an interesting problem. The current bounds use some results of Rademacher average in statistical learning theory \cite{Vap98,Vap13,BBL04,BBL05}. 
However, we find that based on the ideas given by \cite{BBL04,Toi96}, we can develop a new upper bound \emph{without} Rademacher average. 
We also find that this new bound is asymptotically tighter than the existing bounds, i.e., given the chosen samples, as the allowed error probability approaches zero, the new bound is roughly only half of the existing ones. This implies that by using the new bound, a progressive sampling based FI approximation algorithm can reach the guaranteed accuracy with much fewer samples. We also notice that there is no parameter in the new bound that needs to be progressively computed. Hence the sample size that will guarantee the worst error can be precomputed.

Based on the similar idea, we also consider the top-$k$ FI mining problem, which seeks for the $k$ most frequent itemsets in the observed ones. We need to decide when the sampling should stop. The number of the sampled transactions is enough if the worst-case error upper bound is less than the frequency gap between the $k$-th and the $(k+1)$-th most frequent itemsets. Hence we propose a progressive approximation algorithm to address the top-$k$ FI mining problem.

{\bf Our Contributions}. We give a worst-case error upper bound that is asymptotically tighter than the state-of-art bounds, and propose an approximation algorithm which can guarantee the worst-case error upper bound with precomputed sample size. We also give a progressive sampling algorithm to find the top-$k$ most frequent itemsets. Combining with existing methods, our algorithms can approximate the frequent itemsets accurately and efficiently.

The rest of this paper is organized as follows. Section \ref{sec:rw} reviews the related research works. Section \ref{sec:prlm} introduces the notations and preliminaries throughout this paper. Section \ref{sec:refine} gives the worst-case error upper bound without Rademarcher average and compares it with the existing ones. Section \ref{sec:algs} proposes our approximation algorithms based on our upper bounds. Section \ref{sec:eval} gives our evaluation results, which compare our algorithms with the state-of-art.

\section{Related Works}
\label{sec:rw}
Frequent Itemset Mining has been very popular in the communities of information retrieval and data mining \cite{LRU14}. Unsurprisingly, many algorithms that can compute the exact frequencies have been proposed, e.g., A-Priori algorithm \cite{AIS93}, Park-Chen-Yu's algorithm \cite{PCY95}, Multistage algorithms \cite{FSG99}.
However it is very challenging to deal with large scaled data sets with limited main memory. Thus the classical exact algorithms may not fit well in practice. As a result, how to approximate the frequent itemsets by sampling has become interesting, since usually the sample size is much less than the entire data scale. Toivonen \cite{Toi96} was among the first to study sampling on FI approximation, and suggested the first worst-case error bound on frequencies. However his algorithm did not directly use the bound and still needed to parse all the dataset. Thus for scenarios like streams where the size of dataset is unbounded, we cannot use Toivonen's algorithm directly.

Sampling-based frequent itemset approximation has been studied extensively by the researchers. The first works on this problem used heuristic methods to progressively approximate the frequencies \cite{CHS02,CCY05,Parth02}. There were no guarantee on the worst-case error upper bound. To fix this, Riondato and Upfal were the first to propose FI approximation algorithms that could guarantee the worst error bounds by using results of Vapnik-Chervonenkis (VC) dimension \cite{RU12,RU14} and Rademacher average \cite{RU15}. Note that in statistical learning theory VC dimension and Rademacher average are usually used to address the worst-case error upper bound for \emph{infinite case}, i.e., the number of possible functions in the learning model is infinite \cite{BBL04}. However in the case of FI mining, since there are only \emph{finite} itemsets, it is possible to develop bounds without VC dimension or Rademacher average \cite{BBL04}. In this paper we apply this idea on FI mining problem. Riondato et al. also considered using parallelism in FI mining \cite{RDF12}, which is an orthogonal topic to sampling-based FI approximation.

In practice we are often only interested in the most frequent itemsets. Thus top-$k$ FI mining is a popular research topic with many research works \cite{PRU10,SW02,RU15,RV14}. Another interesting question is to find all itemsets with frequencies larger than a threshold. Savasere, Omiecinski, and Navathe \cite{SON95} give an two-pass algorithm (called SON algorithm) that can find the exact solutions. We will use SON algorithm to significantly reduce the number of itemsets to be observed, and then apply our algorithms to approximate the frequencies and select the top $k$ ones. Also Toivonen’s Algorithm \cite{Toi96} is an alternative way to find the most frequent itemsets given a threshold.

\section{Preliminaries}
\label{sec:prlm}
\subsection{Frequency of Itemset}
\newcommand{\I}{\mathcal{I}}
\newcommand{\D}{\mathcal{D}}
In this paper we use the notations and definitions from Riondato and Upfal's pioneering work \cite{RU15}. 
Let $\I$ be the set of items. A transaction $\tau$ is a subset of $\I$ (i.e., $\tau \subseteq \I$).
An itemset $A$ is a set of items that appear together in a transaction $\tau$, i.e., $A \subseteq \tau$. Clearly any itemset is also a subset of $\I$. 
Let transactional dataset $\D$ be the set of all the transactions. In this paper we always assume $\D$ is a finite set. Denote by $T_\D(A)$ the set of all the transactions in $\D$ that contain the itemset $A$. $T_\D(A)$ is also known as the support set of $A$ in $\D$.
If $\D$ is a finite set, we can define the frequency of itemset $A$ in $\D$ as the fraction of transactions in $\D$ that contain $A$.
$$f_\D(A) = |T_\D(A)|/|\D|.$$
Clearly $0 \leq f_\D(A) \leq 1$ for any $A \subseteq \I$.

The goal of our sampling algorithm is to approximate $f_\D(A)$ given an itemset $A$ as accurately as possible.

\subsection{Approximation Algorithms}
\newcommand{\Smp}{\mathcal{S}}
An $(\epsilon,\delta)$-approximation algorithm of the frequencies $f_\D(\cdot)$ takes as input all the items $\I$ and outputs a sampled average $f_\Smp(A)$ for each $A\subseteq\I$ such that with probability at least $1-\delta$,
$$\max_{A\subseteq\I}|f_\D(A) - f_\Smp(A)| \leq \epsilon.$$
We often use progressive sampling \cite{RU15,RU16}, i.e., to keep taking more samples until a stopping condition is reached. A stopping condition usually takes the form $\Delta(n, \delta) \leq \epsilon$, where $n$ is the number of samples that have been taken, and $\Delta$ is an upper bound of the worst approximation error given by statistical learning theory. Note that $\Delta$ is usually a function of $n$ and $\delta$.

There is a variant called top-$k$ approximation, which returns the $k$ most frequent itemsets among the observed ones based on the approximated frequencies. This is quite popular in practice since we are often only interested in the most common itemsets. 

\subsection{Risk Bounds}
\label{sec:rb}
\newcommand{\R}{\mathcal{R}}
We briefly review some risk bounds in statistical learning theory \cite{BBL05} with the background of frequent itemsets mining. 

For each itemset $A\subseteq\I$, define the indicator function $\phi_A : 2^\I \to \{0, 1\}$ as follows.
$$\phi_A(\tau) = \begin{cases}
1 & \textrm{if $A\subseteq \tau$}\\
0 & \textrm{otherwise}\\
\end{cases},\quad
\tau\subseteq\I.$$
Clearly, the frequency $f_\D(A)$ equals to the \emph{true} average of $\phi_A(\tau)$ where $\tau$ goes over all the transactions in $\D$.
$$f_\D(A) = \frac{1}{|\D|} \sum_{\tau\in\D} \phi_A(\tau).$$
Similarly let $\Smp$ be the set of the sampled transactions. Then the \emph{sampled} average of $\phi_A(\tau)$ can be defined as
$$f_\Smp(A) = \frac{1}{|\Smp|} \sum_{\tau\in\Smp} \phi_A(\tau).$$
Clearly $f_\Smp(A)$ is the frequency of $A$ appearing in the sampled transactions $\Smp$.

Assume $|\Smp| = n$. For each transaction $\tau_i \in \Smp$, let $\sigma_i$ be a Rademacher random variable taking value from $\{-1, 1\}$ with uniform probability distribution. The $\sigma_i$'s are independent. Assuming $\I$ is finite, we define the sample conditional Rademacher average as follows.
$$\R_\Smp = \mathbb{E}_\sigma \left[\max_{A\subseteq\I}\frac{1}{n}\sum_{i=1}^n \sigma_i\phi_A(\tau_i)\right],$$
where $\mathbb{E}_\sigma$ denotes the expectation taken over all the random variables $\sigma_i$'s, conditionally on the sample $\Smp$. 

The following theorem tells us that Rademacher average can be used to upper bound the approximation error, even for the worst case.

\begin{theorem}
\label{thm:old}
(Theorem 3.2, \cite{BBL05}) For any $\delta>0$, with probability at least $1-\delta$,
$$\max_{A\subseteq\I} |f_\D(A) - f_\Smp(A)|\leq 2\R_\Smp + \sqrt{\frac{2\log(2/\delta)}{n}}.$$
\end{theorem}

If we want to use the upper bound given in Theorem \ref{thm:old} in an approximation algorithm, we still need to upper bound the $\R_\Smp$. A classical result is given by Massart \cite{Mas00}.

\begin{theorem}
\label{thm:massart}
(Lemma 5.2, \cite{Mas00}) Let $\ell = \max_{A\subseteq\I} [\sum_{i=1}^n\phi_A(\tau_i)^2]^{1/2}$ where each $\tau_i\in\Smp$. Then
$$\R_\Smp \leq \frac{\ell}{n}\sqrt{2\log N},$$
where $N = 2^{|\I|}$ and $n = |\Smp|$.
\end{theorem}
Hence we have the following stopping condition for an $(\epsilon,\delta)$-approximation sampling algorithm.
$$\Delta_1 := \frac{2\ell}{n}\sqrt{2\log N} + \sqrt{\frac{2\log(2/\delta)}{n}} \leq \epsilon.$$
However for many applications the above bound is not tight enough \cite{RU15,RU16}. In the next section we will first review the state-of-art bound on the worst approximation error, and then give an asymptotically tighter bound.

\section{Refining the Upper Bound}
\label{sec:refine}
The reason why the bound given in the previous section is often not tight enough in practice is that the $\ell$ defined in Theorem \ref{thm:massart} can be quite large. Suppose there is an itemset $A$ that almost always appears in every transaction in $\D$. Then no matter which sample the algorithm chooses, $\ell$ is roughly $\sqrt{n}$. For $\delta=0.01$, $N = 2^{1000}$, even 100,000 samples are taken, the upper bound is still larger than 0.15. For many applications such an upper bound cannot be acceptable and thus we need to take more samples. Clearly if the upper bound is tighter, a lot of samples can be saved.

\subsection{A Brief Review on the Existing Results}
Riondato and Upfal \cite{RU15} attempted to give a tighter bound of the Rademacher average $\R_\Smp$. 

\begin{theorem}
\label{thm:ru}
(Theorem 3, \cite{RU15}, revised) Let $w : \mathbb{R}^+ \to \mathbb{R}^+$ be the function defined as
$$w(s) = \frac{1}{s}\log \sum_{A\subseteq\I}\exp\left(\frac{s^2 \sum_{i=1}^n \phi_A(\tau_i)^2}{2n^2}\right).$$
Then $\R_\Smp \leq \min_{s>0} w(s)$.
\end{theorem}

{\bf Remark}. Note that in Theorem \ref{thm:ru}, the summation in $w(s)$ takes \emph{exactly} $2^{|\I|}$ terms. However in the original version in \cite{RU15}, the authors claimed that the summation could take much less than $2^{|\I|}$ terms. We argue that there is a gap between these two versions. Based on the proof given in \cite{RU15}, one can reach the inequality as follows.
\begin{equation}
\label{eqn:ru}\exp(s\R_\Smp) \leq \sum_{A\subseteq\I}\exp\left(\frac{s^2\sum_{i=1}^n \phi_A(\tau_i)^2}{2n^2}\right).
\end{equation}
Note that on the right hand side, each term in the summation is no less than 1. Hence when taking the logarithm on both sides and dividing by $s$, each of the $2^{|\I|}$ terms cannot be eliminated. Thus the range of the summation cannot be compressed.

Formally, suppose there is a set $\mathcal{V}\subseteq 2^\I$, where $2^\I$ denotes the power set of $\I$, such that
$$\alpha(s) := \sum_{A\in 2^\I}\exp\left(\frac{s^2\sum_{i=1}^n \phi_A(\tau_i)^2}{2n^2}\right) \leq \sum_{A\in \mathcal{V}}\exp\left(\frac{s^2\sum_{i=1}^n \phi_A(\tau_i)^2}{2n^2}\right) :=\beta(s).$$
We take the limits as $s$ approaches 0.
$$2^{|\I|}=\lim_{s\to 0}\alpha(s) \leq \lim_{s\to 0}\beta(s) = |\mathcal{V}|.$$
Hence $\mathcal{V} = 2^\I$. This implies any summation over only a part of $2^\I$ must be less than the summation over all of $2^\I$. Thus one cannot use Inequality (\ref{eqn:ru}) to reach Theorem 3 in \cite{RU15}.

\subsection{Tighter Bound Without Rademacher Average}
In statistical learning theory, the upper bound given by Theorem \ref{thm:old} is for the general case, i.e., the set of itemsets can be infinite or finite. However, for frequent itemsets mining, the number of itemsets is always finite (at most $2^{|\I|}$). Given this assumption, can we establish any upper bound without using the Rademacher average? Following the similar lines given by Boucheron, Bousquet and Lugosi \cite{BBL04} and Toivonen \cite{Toi96}, we can give a positive answer.

For any $\epsilon > 0$,
$$\begin{aligned}
& \Pr[\max_{A\subseteq\I}|f_\D(A)-f_\Smp(A)|>\epsilon] \\
= & \Pr[\exists A\subseteq\I, f_\D(A)-f_\Smp(A)>\epsilon \vee f_\D(A)-f_\Smp(A)<-\epsilon] \\
\leq & \Pr[\exists A\subseteq\I, f_\D(A)-f_\Smp(A)>\epsilon] +\Pr[\exists A\subseteq\I, f_\D(A)-f_\Smp(A)<-\epsilon] && \textrm{(union bound)}\\
\leq & \sum_{A\subseteq\I} \Pr[f_\D(A)-f_\Smp(A)>\epsilon] + \sum_{A\subseteq\I} \Pr[f_\Smp(A)-f_\D(A)>\epsilon] && \textrm{(union bound)}.
\end{aligned}$$
Recall Hoeffding's inequalities \cite{H63}. Let $X_1, \cdots, X_n$ be independent random variables bounded by the intervals $[a_i, b_i]$. Define the sampled average of them as
$$\overline{X} = \frac{1}{n}\sum_{i=1}^n X_i.$$
Then for any $t>0$,
$$\Pr[\overline{X} - \mathbb{E}[\overline{X}] > t] \leq \exp\left(-\frac{2n^2t^2}{\sum_{i=1}^n (b_i-a_i)^2}\right),$$
and
$$\Pr[\mathbb{E}[\overline{X}] - \overline{X} > t] \leq \exp\left(-\frac{2n^2t^2}{\sum_{i=1}^n (b_i-a_i)^2}\right).$$
Note that if we set $X_i = \phi_A(\tau_i)$, then $X_i$'s are independent since $\tau_i$'s are independent, and thus $f_\D(A) = \mathbb{E}[\overline{X}]$ and $f_\Smp(A) = \overline{X}$. Based on Hoeffding's inequalities, 
$$\Pr[f_\D(A) - f_\Smp(A) > \epsilon] \leq \exp\left(-2n\epsilon^2\right),$$
and
$$\Pr[f_\Smp(A) - f_\D(A) > \epsilon] \leq \exp\left(-2n\epsilon^2\right).$$
Putting the above results together, we have
$$\begin{aligned}
& \Pr[\max_{A\subseteq\I}|f_\D(A)-f_\Smp(A)|>\epsilon] \\
\leq & 2\sum_{A\subseteq\I} \exp\left(-2n\epsilon^2\right) \\
= & 2N \exp\left(-2n\epsilon^2\right),
\end{aligned}$$
where $N = 2^{|\I|}$. 
Equivalently for any $\delta>0$, with probability at least $1-\delta$,
$$\max_{A\subseteq\I}|f_\D(A)-f_\Smp(A)| \leq \sqrt{\frac{\log(2N) + \log(1/\delta)}{2n}} =: \Delta_2.$$
Note that the above bound $\Delta_2$ is very similar to the result in Section 3.4, \cite{BBL04}. Now the bound $\Delta_2$ can generate a new stopping condition for an approximation algorithm.

Recall the classical upper bound given in Section \ref{sec:rb}.
$$\Delta_1 := \frac{2\ell}{n}\sqrt{2\log N} + \sqrt{\frac{2\log(2/\delta)}{n}}.$$
Clearly $\lim_{\delta\to 0}\Delta_1/\Delta_2 = 2$, i.e., when $\delta$ is very small, the bound $\Delta_1$ is roughly twice of $\Delta_2$ given the sample size $n$.
This assures us that the bound $\Delta_2$ is highly competitive.

Theorem \ref{thm:ru} can give another upper bound on the worst approximation error. However, since the number of terms in the summation grows exponentially on $|\I|$, to find the minimum is computationally infeasible. Furthermore, even if the minimum $w(s^*)$ is found, let $\Delta_1'$ be the upper bound of this variant defined as 
$$\Delta_1' := w(s^*) + \sqrt{\frac{2\log(2/\delta)}{n}}.$$
By fixing the sample $\Smp$, we still have $\lim_{\delta\to 0}\Delta_1'/\Delta_2 = 2$. For small $\delta$, the bound without Rademacher average still outperforms the existing ones.

\section{Our Frequent Itemset Approximation Algorithm}
\label{sec:algs}
\subsection{Approximating with Precomputed Sample Size}
\label{sec:nonprg}
We observe the bound given in the previous section:
$$\Delta_2 := \sqrt{\frac{\log(2N) + \log(1/\delta)}{2n}},$$
where $N = 2^{|\I|}$.
The upper bound $\Delta_2$ can be treated as a function of allowed error probability $\delta$, sample size $n$ and $N = 2^{|\I|}$, all of which are already given. A good news is that there is no parameter that needs to be progressively computed (e.g., $\ell$ in $\Delta_1$). Thus to guarantee an worst approximation error at most $\epsilon$, we only need to make sure $\Delta_2 \leq \epsilon$. By solving it we have
$$n \geq \frac{1}{2\epsilon^2}(\log (2N) + \log (1/\delta)).$$
Note that this sampling bound agrees with Toivonen's result (Corollary 2 in \cite{Toi96}).

Hence an $(\epsilon,\delta)$-approximation algorithm takes a very simple form. We first consider a brute-force algorithm to approximate frequencies for \emph{all} the itemsets. Note that since the number of subsets (itemsets) in $\I$ is exponential (i.e., $2^{|\I|}$), the brute-force algorithm is not efficient.

\framebox{
\begin{minipage}{.9\textwidth}
\underline{\bf Frequent Itemsets Approximation Algorithm (brute-force for all itemsets)}

{\bf Input}: items $\I$, transactional dataset $\D\subseteq 2^\I$, $\epsilon>0$, $\delta>0$.

{\bf Output}: approximated frequencies $\hat{f}_\D(A)$ for each $A\subseteq\I$ s.t. with probability at least $1-\delta$, $|\hat{f}_\D(A)|\leq \epsilon$ for any $A\subseteq\I$.

\begin{enumerate}
\item $n \gets \lceil\frac{1}{2\epsilon^2}(\log(2^{|\I|+1}) + \log (1/\delta))\rceil$.
\item $\Smp \gets \emptyset$.
\item If $n\geq |\D|$, $\Smp \gets \D$; otherwise, choose $n$ itemsets in $\D$ at uniformly random and add them to $\Smp$.
\item Label the transactions in $\Smp$: $\Smp = \{\tau_1,\cdots,\tau_n\}$.
\item For each $A\subseteq\I$, compute $\hat{f}_\D(A) \gets \frac{1}{n}\sum_{i=1}^n \phi_A(\tau_i)$.
\item Return all the $\hat{f}_\D(A)$'s for $A\subseteq\I$.
\end{enumerate}
\end{minipage}}

Since the brute-force algorithm above is computationally infeasible when $|\I|$ is large, in practice, we often only consider the frequencies of a few itemsets, e.g., most popular pairs of complementary goods, influential coauthoring in a community, etc. For this case, we do not have to consider the itemsets, which do not appear frequently enough. 
\newcommand{\Ob}{\mathbf{Ob}}
Denote by $\Ob$ the set of the itemsets to be observed. Then the worst approximation error is defined as the maximum error on every itemset in $\Ob$. By the same reasoning in the derivation of $\Delta_2$, we have the adjusted new bound:
$$\Delta_2' := \sqrt{\frac{\log(2|\Ob|) + \log(1/\delta)}{2n}}.$$
Since $\Ob$ is a subset of $2^{\I}$, this bound $\Delta_2'$ is tighter than $\Delta_2$. Note that Toivonen (Corollary 2, \cite{Toi96}) also found a similar result as our bound here. The approximation algorithm will also be revised as follows.

\framebox{
\begin{minipage}{.9\textwidth}
\underline{\bf Frequent Itemsets Approximation Algorithm with Observed Itemsets}

{\bf Input}: all the items $\I$, the observed itemsets $\Ob\subseteq 2^\I$, transactional dataset $\D\subseteq 2^\I$, $\epsilon>0$, $\delta>0$.

{\bf Output}: approximated frequencies $\hat{f}_\D(A)$ for each $A\in\Ob$ s.t. with probability at least $1-\delta$, $|\hat{f}_\D(A)|\leq \epsilon$ for any $A\in\Ob$.

\begin{enumerate}
\item $n \gets \lceil\frac{1}{2\epsilon^2}(\log(2|\Ob|) + \log (1/\delta))\rceil$.
\item $\Smp \gets \emptyset$.
\item If $n\geq |\D|$, $\Smp \gets \D$; otherwise, choose $n$ itemsets in $\D$ at uniformly random and add them to $\Smp$.
\item Label the transactions in $\Smp$: $\Smp = \{\tau_1,\cdots,\tau_n\}$.
\item For each $A\in\Ob$, compute $\hat{f}_\D(A) \gets \frac{1}{n}\sum_{i=1}^n \phi_A(\tau_i)$.
\item Return all the $\hat{f}_\D(A)$'s for $A\in\Ob$.
\end{enumerate}
\end{minipage}}

Note that we do not have to estimate for any itemset which is out of the observed ones $\Ob$. Also we need the size of $\Ob$ to be as small as possible. Depending on the practical requirements, the choice of $\Ob$ can vary a lot. We will give a SON-based idea in the next section. However many other methods can be tried, e.g., most potentially frequent itemsets can be suggested by the users' experience or historic records.

\subsection{Approximating Top-$k$ Frequent Itemsets}
\label{sec:prg}
In practice we often need to find out the top-$k$ frequent itemsets among the given candidates $\Ob$. We can slightly revise the algorithm given in the previous section to approximate the $k$ most frequent itemsets. A new problem here is how to give the stopping condition. Note that if we only need the top-$k$ frequent itemsets, then our approximation can stop when the members of top-$k$ FIs are fixed with high probability (i.e., at least $1-\delta$). In particular, if with probability at least $1-\delta$, the true frequency of any itemset will not surpass the middle point of the $k$-th and $(k+1)$-th largest approximated frequencies, the $k$ itemsets with largest approximated frequencies should probably be the correct top $k$ ones. By Hoeffding's inequality and union bounds, given the approximate frequencies with $n$ samples, the probability $p$ that there exists an itemset, whose approximated frequency and true frequency are on the different sides of the middle point of the $k$-th and $(k+1)$-th largest approximated frequencies, can be upper bounded as follows: 
$$\begin{aligned}
p&=\Pr[\bigvee_{A\in\Ob} (f_\D(A) < m < \hat{f}_\D(A)) \vee (f_\D(A) > m > \hat{f}_\D(A)) ] \\
&\leq \sum_{A\in\Ob} \exp\left(-2n(\hat{f}_\D(A) - m)^2\right),
\end{aligned}$$
where $m$ is the frequency middle point as described above. 
Hence we can let the sampling stop when the upper bound of $p$ is less than $\delta$. 
Combining these ideas, a progressive sampling approximation algorithm can be given as follows:

\framebox{
\begin{minipage}{.9\textwidth}
\underline{\bf Top-$k$ Frequent Itemsets Approximation Algorithm with Observed Itemsets}

{\bf Input}: all the items $\I$, the observed itemsets $\Ob\subseteq 2^\I$, transactional dataset $\D\subseteq 2^\I$, sampling increase $\Delta n$, $k>0$, $\epsilon>0$, $\delta>0$.

{\bf Output}: $k$ itemsets among $\Ob$ that have the highest approximate frequencies s.t. with probability at least $1-\delta$, the approximate frequencies have worst-case error less than $\epsilon$.

\begin{enumerate}
\item $n\gets 0$. $\hat{f}_\D(A) \gets 0$ for any $A\in\Ob$.
\item $\Smp \gets \emptyset$, $N \gets \min\{|\D|,\lceil\frac{1}{2\epsilon^2}(\log(2|\Ob|) + \log (1/\delta))\rceil\}$.
\item Choose $\Delta n$ itemsets from $\D$ at uniformly random, and denote by $\Delta \Smp$ the chosen $\Delta n$ itemsets.
\item Label the transactions in $\Delta\Smp$: $\Delta\Smp = \{\tau_1,\cdots,\tau_{\Delta n}\}$.
\item For each $A\in\Ob$, compute $\hat{f}_\D(A) \gets \frac{1}{n+\Delta n}\left(n\cdot\hat{f}_\D(A)+\sum_{i=1}^{\Delta n}\phi_A(\tau_i)\right)$.
\item $n \gets n+\Delta n$. $\Smp \gets \Smp \cup \Delta \Smp$
\item Let $\hat{f}[k]$ and $\hat{f}[k+1]$ be the $k$-th and $(k+1)$-th largest approximate frequencies in $\Ob$ respectively. Compute their middle point $m \gets \frac{1}{2}(\hat{f}[k] + \hat{f}[k+1])$.
\item If $n>N$, then return the approximate top-$k$ frequent itemsets $\hat{f}_\D(A)$ for $A\in\Ob$. Otherwise, go to the next step.
\item If the following stopping condition is satisfied: 
$$\sum_{A\in\Ob} \exp\left(-2n(\hat{f}_\D(A) - m)^2\right) < \delta,$$ 
then return the approximate top $k$ frequent itemsets in $\Ob$. Otherwise, go back to Step 3.
\end{enumerate}
\end{minipage}
}

Note that in out top-$k$ approximation algorithm, the stopping condition depends on the $k$-th and $(k+1)$-th largest frequencies. If these two frequencies tie, it is likely that many samples will be needed since we cannot distinguish them based on approximated frequencies. Hence we require the number of samples should not exceed $N$, the number of samples that can guarantee the $(\epsilon, \delta)$-approximation. If more than $N$ samples are needed, we can assume that the $k$-th and $(k+1)$-th largest frequencies tie or are very close. Then we directly output the approximated top FIs, since further computations to distinguish the very close FIs on the boundary are often unnecessary in practice. 

To be efficient, we must ensure the size of $\Ob$ is small enough. One possible way, which is similar to A-Priori algorithm \cite{AIS93}, is given as follows:
\begin{enumerate}
\item We first only consider the itemsets with single item. The item size is usually small enough (i.e., $|\I|$) that can be put in main memory. We approximate their frequencies, and take the threshold $T$ as the $k$-th largest frequency among the single items. Usually people are only interested in small $k$, e.g., 10 to 100, which is much less than $|\I|$.
\item Then we use SON algorithm $\cite{SON95}$ to exactly find the itemsets with frequencies at least $T$. For efficiency, in SON we only consider the itemsets with 2 items, like what \cite{LRU14} did. The reason is that the itemsets of sizes larger than 2 usually have much lower frequencies than pairs. Clearly SON algorithm can find at most $k^2$ candidate itemsets.
\item We use our (top-$k$) approximation algorithms to estimate the frequencies of the candidate itemsets (or select the top-$k$ ones).
\end{enumerate}
Since we only consider frequent pairs, we can also build our scheme on PCY or Multistage algorithms. The major difference between PCY, Multistage and A-Priori is how to fully use the main memory for the passes, in which we select the most frequent items or itemsets. Hence the difference does not affect our sampling. Also we can use Toivonen's algorithm instead of SON to mine frequent itemsets with more than 2 items.

\section{Evaluation}
\label{sec:eval}
In this section we present our evaluation results. 
We try to find the top-$K$ frequent itemsets (in pairs) by two algorithms proposed in this paper:
\begin{itemize}
\item {\bf A-Priori + Precomputed Sample Size}. We first use A-Priori algorithm to find the top-$K$ most frequent items, and then use the algorithm (discussed in Section \ref{sec:nonprg}) to approximate the frequencies of all the pairs formed by the $K$ items. At last we sort the frequencies and find the top-$K$ pairs with the highest approximate frequencies. Note that the sample size can be precomputed.
\item {\bf A-Priori + Progressive Sampling}. This is given in Section \ref{sec:prg}. Note that the sampling is progressive, i.e., there is no precomputed sample size.
\end{itemize}
We compare our algorithms with the state-of-art \cite{RU15}, which is a progressive sampling approximation. For each comparison, our code for each algorithm is similarly organized except that the bounds are different. 
For the performance, we consider time complexity (running time), sample size, and precision/recall.

\subsection{Setup}
We implemented our algorithms by Python 3.4.3 and ran the programs on knot cluster (one DL580 nodes with 4 Intel X7550 eight core processors and 512GB RAM) at UCSB Center for Scientific Computing. To reduce the entire running time in the experiments (it is very time consuming to compute the exact solutions), for each dataset, we selected the first 70 items and then approximate the top 10 frequencies of their pairs. Each approximation was repeated by 10 times and we took the averaged results. By default we chose that the sample size increases by 100 for each round, and $\epsilon = 0.05$, $\delta = 0.01\%$.

\subsection{Datasets}

\begin{table}[!t]
\centering
\begin{tabular}{l | r r}
\specialrule{1pt}{1pt}{1pt}
Name & No. of Transactions & No. of items \\
\hline
accidents & 340183 & 468 \\
chess & 3196 & 75 \\
connect & 67557 & 129 \\
kosarak & 990002 & 41270 \\
mushroom & 8124 & 119 \\
pumsb & 49046 & 7116 \\
pumsb star & 49046 &  7116 \\
retail  & 88162 & 16470 \\
\specialrule{1pt}{1pt}{1pt}

\end{tabular}
\caption{Dataset characteristics}
\label{tab:data}
\end{table}

For consistency, we choose FIMI'03 data repository \cite{GZ04}, the real-world data set from \cite{RU15} (the data repository is available at \texttt{http://fimi.ua.ac.be/data/}). The item and transaction data sizes of the FIMI datasets are given in Table \ref{tab:data}. We will use these data sets to evaluate the samples sizes and worst case errors of our algorithms, and compare our results with the state-of-art algorithms.

\subsection{Worst-case Error Upper Bound Comparison}
\begin{figure}[!t]
\centering
\begin{minipage}{.45\textwidth}
\includegraphics[width=\textwidth]{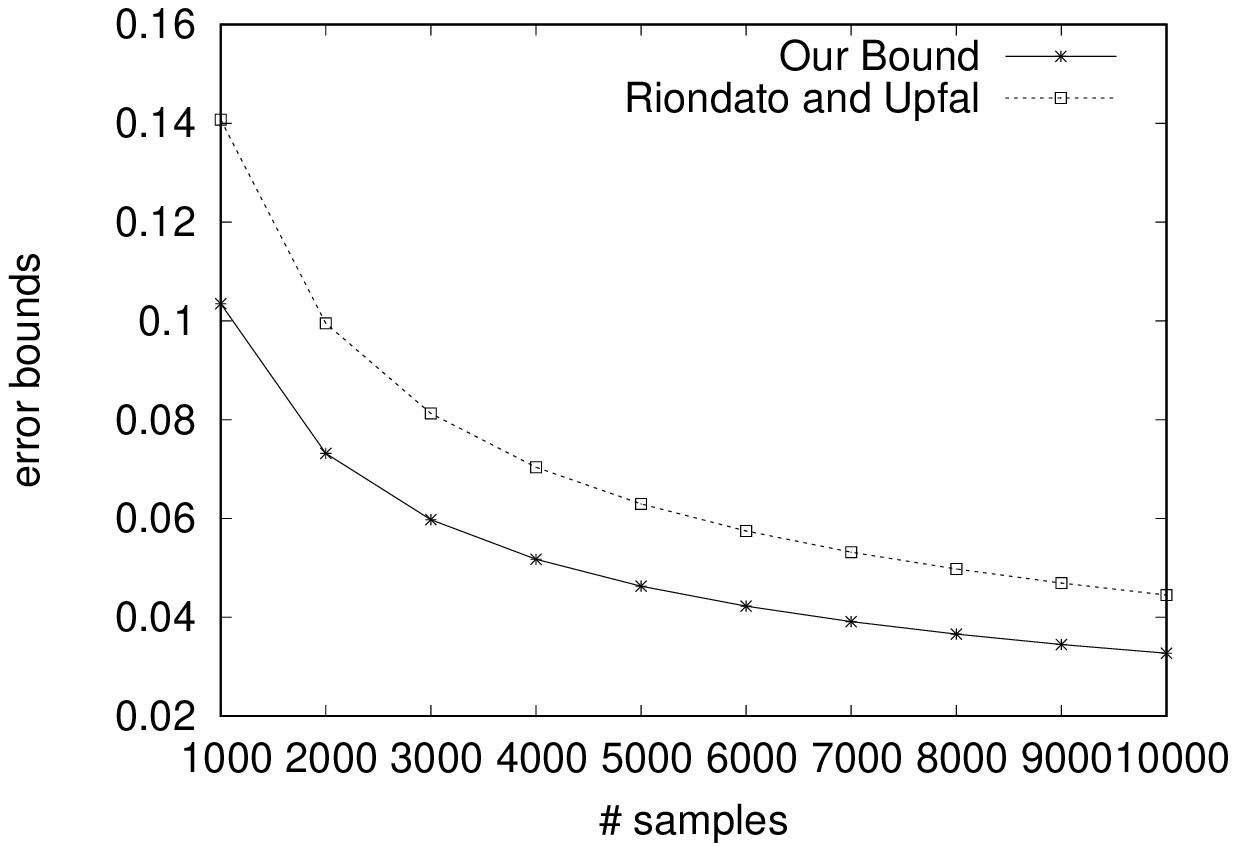}
\caption{$\epsilon = 0.01$, $\delta = 0.0001$.}
\label{fig:b1}
\end{minipage}
\hspace{5pt}
\begin{minipage}{.45\textwidth}
\includegraphics[width=\textwidth]{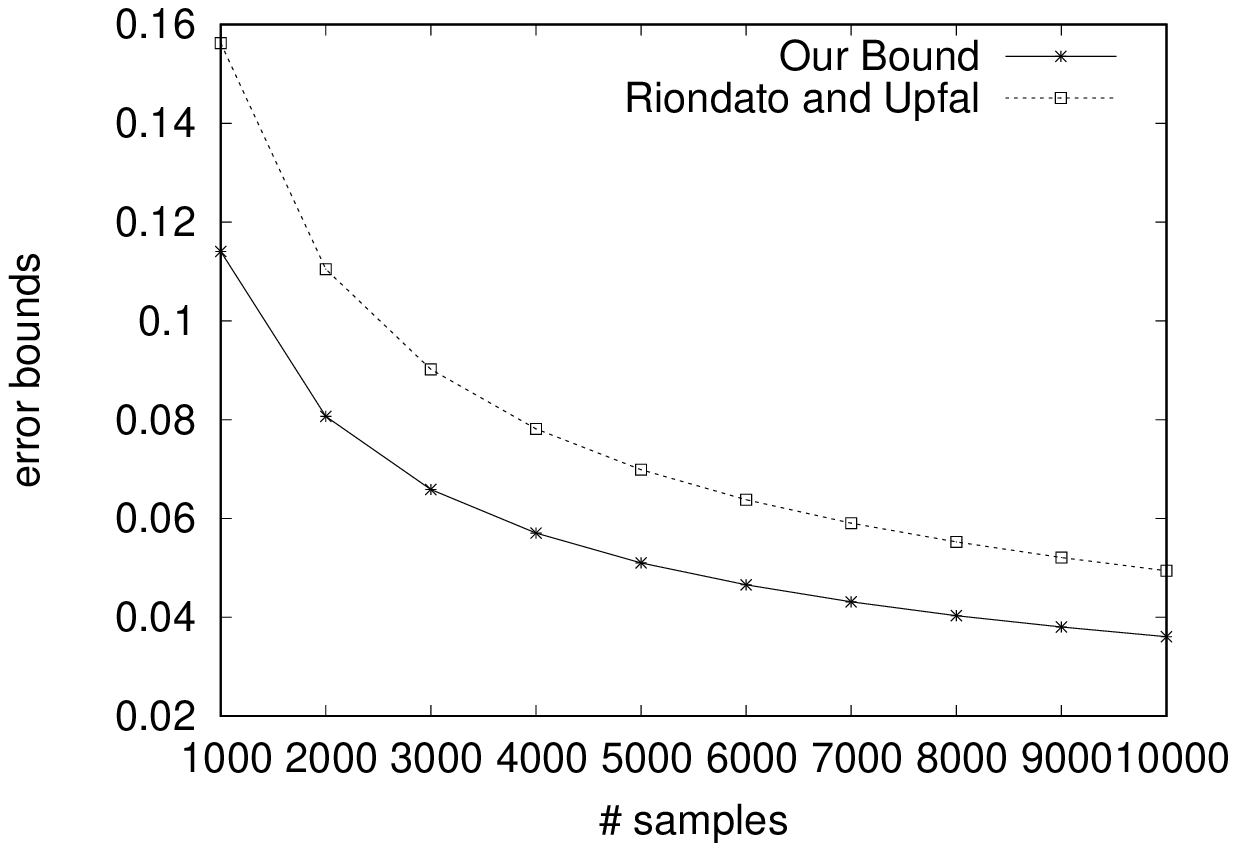}
\caption{$\epsilon = 0.001$, $\delta = 0.00001$.}
\label{fig:b2}
\end{minipage}
\end{figure}

We compare our new worst-case error bound with the state-of-art \cite{RU15}.
Figure \ref{fig:b1} and Figure \ref{fig:b2} give our bounds on worst-case errors and \cite{RU15}'s for different combinations of $\epsilon$, $\delta$ and sample size. 
Clearly our new bound outperforms the state-of-art, implying that to achieve a certain degree of accuracy, compared to \cite{RU15}'s estimation, actually much fewer samples are needed. This key result gives the basic motivation of our algorithms.

\subsection{Comparison Results}
We compare the performance between our methods and the state-of-art \cite{RU15}.
Table \ref{tab:res1} gives the evaluation results for our algorithms discussed in Section \ref{sec:nonprg}. The algorithms try to find the top-100 most frequent itemsets with two items. Our algorithm will first find the 100 most frequent items, and then approximate the frequencies of the pairs between the 100 items. Thus the number of the observed itemsets is $|\Ob| = 100*99/2 = 4950$ for the second pass. Given the default $\epsilon$ and $\delta$, the precomputed sample size $n$ for the second pass is fixed, except the dataset chess, whose transactional size is less than the precomputed sample size. Note that we have significantly confined the observed itemsets, and thus it is efficient to compute the upper bound in \cite{RU15}, i.e., we do not need to consider each subset of $\I$. In the table we have included all the samples taken in both the first and second passes.
By using our new error upper bound, the running time and sample size are significantly reduced compared to \cite{RU15}, while the accuracy is still quite competitive, i.e., all are larger than 90\%. This is natural since \cite{RU15} takes more samples and thus the estimation should be more accurate. In the dataset chess, every transaction is sampled since its data volume is very small. In the dataset connect, our sample size is only 1/20 of \cite{RU15}'s, but the accuracy is almost the same.

\newcommand{\specialcell}[2][c]{
  \begin{tabular}[#1]{@{}c@{}}#2\end{tabular}}
\begin{table}[!t]
\centering
\begin{tabular}{l | r r r | r r r}
\specialrule{1pt}{1pt}{1pt}
Item & \specialcell{Used time (sec)\\(ours)} & \specialcell{Sample size\\ (ours)} & \specialcell{Precision\\(ours)} & \specialcell{Used time (sec)\\\cite{RU15}} & \specialcell{Sample size\\\cite{RU15}} & \specialcell{Precision\\\cite{RU15}} \\
\hline
accidents & 48.78 & 6894 & 97\% & 2683.74 & 387782 & 99\% \\
chess & 21.44 & 3196 & 100\% & 21.86 & 3196 & 100\% \\
connect & 50.34 & 6636 & 98\% & 1097.70 & 132751 & 99\% \\
kosarak & 104.31 & 7657 & 90\% & 11735.33 & 949155 & 98\% \\
mushroom & 43.32 & 6620 & 98\% & 125.56 & 18646 & 99\% \\
pumsb & 193.39 & 7438 & 98\% & 2998.37 & 115939 & 98\% \\
pumsb star & 155.15 & 7438 & 95\% & 2595.17 & 133668 & 99\% \\
retail & 102.93 & 7606 & 91\% & 1909.02 & 128544 & 96\% \\
\specialrule{1pt}{1pt}{1pt}
\end{tabular}
\caption{Results of our approximation algorithms with precomputed sample size and \cite{RU15}.}
\label{tab:res1}
\end{table}

Table \ref{tab:res2} gives the evaluation results for our algorithms discussed in Section \ref{sec:prg}.
Similarly to the above non-progressive version, our progressive method is still quite efficient and accurate. The comparison between ours and \cite{RU15} show that our method is competitive.

\begin{table}[!t]
\centering
\begin{tabular}{l | r r r | r r r}
\specialrule{1pt}{1pt}{1pt}
Item & \specialcell{Used time (sec)\\(ours)} & \specialcell{Sample size\\ (ours)} & \specialcell{Precision\\(ours)} & \specialcell{Used time (sec)\\\cite{RU15}} & \specialcell{Sample size\\\cite{RU15}} & \specialcell{Precision\\\cite{RU15}} \\
\hline
accidents & 18.69 & 3600 & 98\% & 357.37 & 60600 & 99\% \\
chess & 21.96 & 3200 & 98\% & 22.33 & 3200 & 99\% \\
connect & 43.04 & 6700 & 99\% & 748.37 & 110200 & 100\% \\
kosarak & 110.84 & 7700 & 88\% & 1367.20 & 86000 & 97\% \\
mushroom & 46.13 & 6700 & 98\% & 100.68 & 16400 & 98\% \\
pumsb & 170.12 & 7500 & 98\% & 2225.19 & 98200 & 98\% \\
pumsb star & 130.49 & 7500 & 98\% & 1708.80 & 98200 & 98\% \\
retail & 98.24 & 7700 & 94\% & 1152.59 & 83700 & 97\% \\
\specialrule{1pt}{1pt}{1pt}
\end{tabular}
\caption{Comparisons on our progressive approximation algorithms and \cite{RU15}.}
\label{tab:res2}
\end{table}

\section{Conclusion}
We have proposed a new upper bound for the worst-case errors on sampling-based approximate frequent itemsets mining. Our new bound is tighter than the state-of-art result. Based on our new bound, two approximation algorithms have been proposed. We have used real-world datasets to evaluate our results and the performance of our algorithms. The evaluation results have shown that our algorithms are not only competitive but also efficient.

\bibliographystyle{plain}
\bibliography{./cs273}

\end{document}